\documentclass[preprint,showpacs,amsmath,amssymb,superscriptaddress,showkeys,prc,nofootinbib]{revtex4}

\usepackage{graphicx}
\usepackage{dcolumn}
\usepackage{bm}
\usepackage{amsmath,amssymb}

\begin{document}


\title{Correlations between azimuthal anisotropy Fourier harmonics in PbPb collisions at $\sqrt{s_{_{\mathrm{NN}}}}=2.76$~TeV in the HYDJET++ and AMPT models}

\author{M. Dordevic}
\affiliation{Vinca Institute of Nuclear Sciences, University of Belgrade, Belgrade, Serbia}

\author{J. Milosevic}
\email[Corresponding author:]{Jovan.Milosevic@cern.ch}
\affiliation{Vinca Institute of Nuclear Sciences, University of Belgrade, Belgrade, Serbia}
\affiliation{University of Oslo, Department of Physics, Oslo, Norway}

\author{L. Nadderd}
\affiliation{Vinca Institute of Nuclear Sciences, University of Belgrade, Belgrade, Serbia}

\author{M. Stojanovic}
\affiliation{Vinca Institute of Nuclear Sciences, University of Belgrade, Belgrade, Serbia} 

\author{F. Wang}
\affiliation{College of Science, Huzhou University, Huzhou, China}
\affiliation{Department of physics and astronomy, Purdue University, West Lafayette, Indiana, USA}

\author{X. Zhu}
\affiliation{College of Science, Huzhou University, Huzhou, China}

\date{\today}

\begin{abstract}
Correlations between azimuthal anisotropy Fourier harmonics $v_{n}$ ($n = 2, 3, 4$) are studied using the events from PbPb collisions at $\sqrt{s_{_{\mathrm{NN}}}}=2.76$~TeV generated by the HYDJET++ and AMPT models, and compared to the corresponding experimental results obtained by the ATLAS Collaboration. The Fourier harmonics $v_{n}$ are measured over a wide centrality range using the two-particle azimuthal correlation method. The slopes of the $v_{2}$--$v_{3}$ correlation from both models are in a good agreement with the ATLAS data. The HYDJET++ model predicts a stronger slope for the $v_{2}$--$v_{4}$ and $v_{3}$--$v_{4}$ correlations than the ones experimentally measured, while the results from the AMPT model are in a rather good agreement with the experimental results. In contrast to the HYDJET++ predictions, the AMPT model predicts a boomerang-like shape in the structure of the correlations as found in the experimental data.
\end{abstract}



\maketitle

\section{Introduction}

\label{intro}
Quantum Chromodynamics predicts that at sufficiently high energy density partons can no longer be confined inside the nucleons. Indeed, a new state of matter with deconfined partons, called Quark-Gluon-Plasma (QGP), is formed in ultra-relativistic nucleus-nucleus collisions~\cite{Shuryak,Busza}. The QGP undergoes a collective expansion which can be described by relativistic hydrodynamics. The initial geometry of the colliding nuclei creates anisotropic pressure gradients in the transverse plane perpendicular to the beam direction. As a consequence, such initial spatial anisotropy is converted into momentum anisotropy observable in the final state as a preferential emission of particles in a certain azimuthal direction. The anisotropic flow can be studied by Fourier decomposition of the emitted hadron yield distribution in azimuthal angle $\phi$~\cite{Ollitrault:1993ba,Voloshin:1994mz,Poskanzer:1998yz}
\begin{equation}
\label{F1}
\frac{dN}{d\phi} \propto 1+2\sum_{n}v_{n}\cos[n(\phi - \Phi_{n})].
\end{equation}
Here, Fourier coefficient $v_{n}$ represents the magnitude of the azimuthal anisotropy measured with respect to the n-th order harmonic plane angle $\Phi_{n}$. The angle $\Phi_{n}$ can be reconstructed from the emitted particle distribution itself. The elliptic flow $v_{2}$ is the most studied anisotropy. The $\Phi_{2}$ which corresponds to the $v_{2}$ is correlated with the participant plane spanned by the beam direction and the shorter axis of the roughly lenticular shape of the nuclear overlap region. The initial-state fluctuations in the positions of nucleons induce higher-order deformations, and thus higher order Fourier harmonics ($v_{n}$, $n \ge $3 in Eq.~(\ref{F1})) are present. Higher-order Fourier harmonics are measured with respect to the corresponding harmonic plane angles $\Phi_{n}$~\cite{Alver:2010gr}. The collective behavior of the QGP has been studied using the azimuthal anisotropy of emitted particles detected in experiments at the Relativistic Heavy Ion Collider (RHIC)~\cite{Back:2002gz,Ackermann:2000tr,Adcox:2002ms} and the Large Hadron Collider (LHC)~\cite{Aamodt:2010pa,ALICE:2011ab,Abelev:2014pua,Adam:2016izf,ATLAS:2011ah,ATLAS:2012at,Aad:2013xma,Chatrchyan:2012wg,Chatrchyan:2012ta,Chatrchyan:2013kba,CMS:2013bza,Khachatryan:2015oea}.

One of the experimental methods used to determine the $v_{n}$ coefficients is based on two-particle azimuthal correlations~\cite{Wang:1991qh}. These correlations can also be Fourier decomposed into
\begin{equation}
\label{F2}
\frac{dN^{pair}}{d\Delta\phi} \propto 1+2\sum_{n}V_{n\Delta}\cos(n\Delta\phi),
\end{equation}
where $\Delta\phi$ is the relative azimuthal angle of the particle pair. Assuming the factorization,
 the two-particle Fourier coefficient $V_{n\Delta}$ is a product of the single-particle anisotropies of the particle pair. The $v_{n}$ anisotropy can then be extracted by
\begin{equation}
\label{v_n}
 v_{n} = \sqrt{V_{n\Delta}},
\end{equation}
with the two particles in the pair correlation belonging to the same particle group. For each event from a given centrality\footnote{The centrality of a nucleus-nucleus collision is defined as a fraction of the total inelastic nucleus-nucleus cross section, with 0\% denoting the most central collisions.} class a two-dimensional two-particle correlation is constructed as a function of $\Delta\phi$ and relative pseudorapidity $\Delta\eta$. Typically, particles with 0.5 $< p_{T} <$ 2~GeV/c and $|\eta| <$~2.5 are used as we adopt in this study. In order to remove short range correlations only pairs with $|\Delta\eta| \ge$ 2 are taken into account. One-dimensional correlation function in $\Delta\phi$ is built by projecting the two-dimensional correlation function onto the $\Delta\phi$ axis and then decomposed using the Eq.~(\ref{F2}).

This paper is organized as follows. The basic features of the HYDJET++ model~\cite{Lokhtin:2008xi} and the AMPT model~\cite{Lin:2004en} are described in Sect.~2. Approximately 1M PbPb collision events at $\sqrt{s_{_{\mathrm{NN}}}}=2.76$~TeV each are simulated using the HYDJET++ (version 2.3) and AMPT (version 1.25-2.25) models. The results and discussions are given in Sect.~3. The results are presented over a wide range of centralities going from ultra central (0-5\% centrality) to peripheral (65-70\% centrality) PbPb collisions. A summary is given in Sect.~4.

\section{HYDJET++ and AMPT model}
\label{HYDJET}
The Monte Carlo HYDJET++ and AMPT models simulate relativistic nucleus-nucleus collisions. HYDJET++ consists of two components which simulate soft and hard processes. The soft part governs the ideal hydrodynamical evolution of the system while the hard part provides multiparton fragmentation. The hard part of the HYDJET++ consists of PYTHIA~\cite{Sjostrand:2006za} and PYQUEN~\cite{Lokhtin:2005px} event generators which simulate initial parton-parton collisions, radiative energy loss of partons and parton hadronization. It also takes into account jet quenching effects within the formed medium. The minimum transverse momentum transfer $p^{min}_{T}$ of a parton-parton scattering determines whether it contributes to the soft or the hard part. In the soft part of the HYDJET++ model, the elliptic flow magnitude is governed by the spatial anisotropy $\epsilon(b)$ which, at a given impact parameter $b$, represents the elliptic anisotropy of the hydrodynamics hypersurface at the freeze-out, and by the momentum anisotropy $\delta(b)$ which is the modulation of flow velocity profile (see Eq.~(34) in~\cite{Lokhtin:2008xi}). In the low-$p_{T}$ range ($\le$~2 GeV/c), the elliptic flow is mainly determined by the internal pressure gradients developed within the expanding fireball during the initial phase of the collision, and is sensitive to the $\epsilon$ and $\delta$ parameters. As each fluid cell carries a certain momentum, at the freeze-out the spatial anisotropy is transformed into the momentum anisotropy. In order to extend the HYDJET++ model to the triangular $v_{3}$ flow, additionally the third-order spatial ($\epsilon_{3}(b)$) and momentum ($\rho(b)$) anisotropy parameters are introduced. Simulation of the events can be performed under several configurations. The most realistic one, 'flow+quenched jets', which includes both hydrodynamics expansion and quenched jets is used in this analysis. The details of the model can be found in the HYDJET++ manual~\cite{Lokhtin:2008xi}.

A multi-phase transport (AMPT) model~\cite{Lin:2004en} consists of several parts: the HIJING model~\cite{Wang:1991hta} which generates semi-hard minijet partons and soft strings; string melting which converts strings into partons; Zhang's parton cascade (ZPC)~\cite{Zhang:1997ej} which simulates the interactions among partons; the Lund string fragmentation~\cite{Lund,Anderson} as implemented in JET-SET/PYTHIA~\cite{Sjostrand} to convert the excited strings into hadrons or a simple quark coalescence model to convert partons into hadrons in the case of string melting; interactions among hadrons are described by the extended relativistic transport model. In this analysis, the default value of the parton cross section of 3~mb is used in ZPC. The running coupling $\alpha_{s}$ and the screen mass $\mu$ values of 0.4714 and 3.2264~fm$^{-1}$ are used, respectively~\cite{Wei:2018xpm}. This, together with the assumed initial temperature of 468~MeV, gives an effective specific shear viscosity $\eta/s$ of 0.137 based on Eq.(4) in~\cite{Wei:2018xpm}. Within the model, the anisotropies of different orders are developed due to the initial state eccentricities. It was shown in~\cite{Zhou:2015eya} that in peripheral collisions the efficiency of converting the initial eccentricities into final momentum anisotropies decreases because of the reduced amount of interactions in the small system. More details about the AMPT model can be found in~\cite{Lin:2004en}.

\section{Results}
\label{Res}
\begin{figure}
  \includegraphics[width=0.7\textwidth]{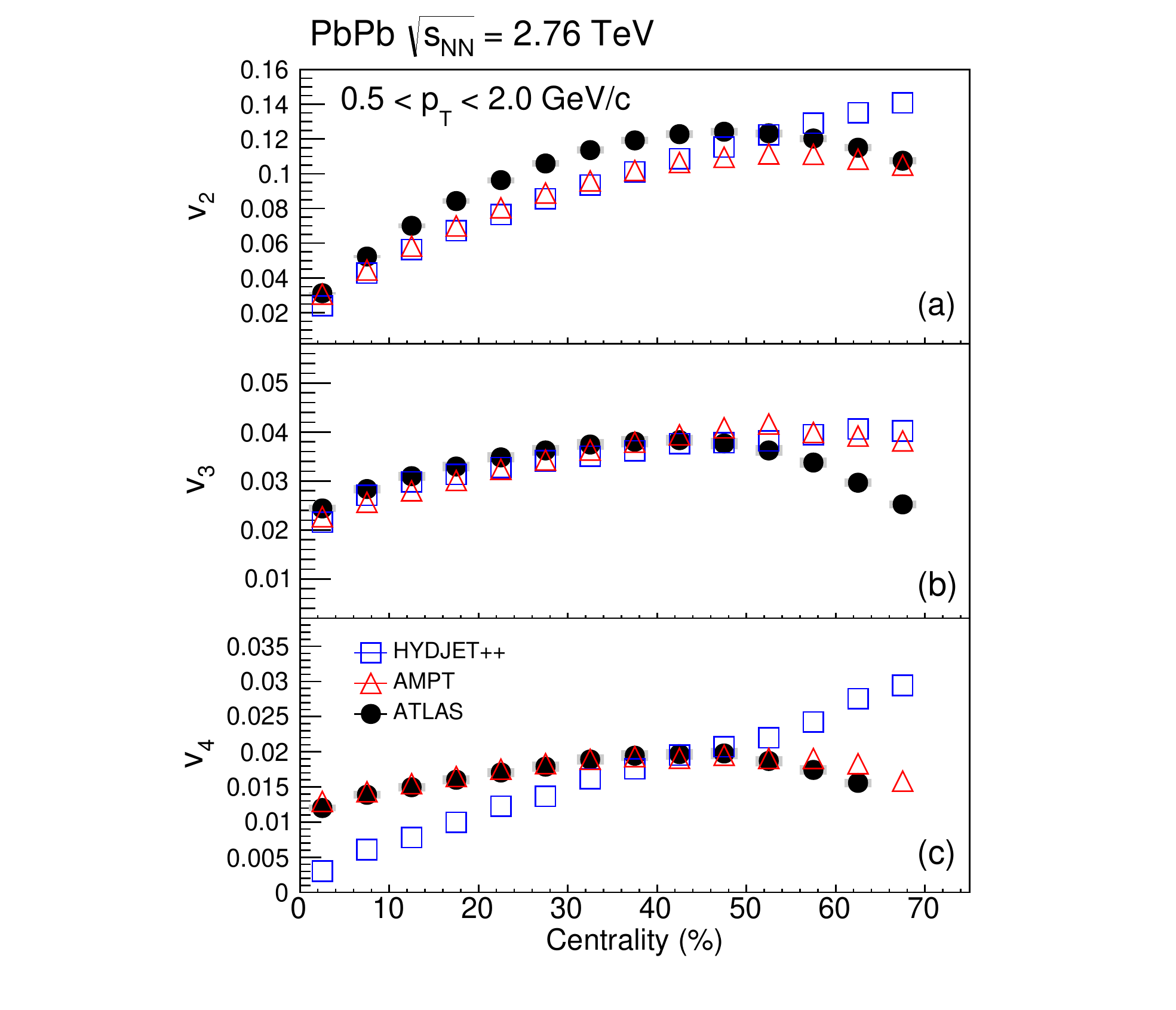}
  \caption{\label{fig:1} (Color online) The centrality dependence of the $v_{2}$ (upper panel), $v_{3}$ (middle panel) and $v_{4}$ (bottom panel) from the 0.5 $< p_{T} <$ 2~GeV/c interval in PbPb collisions at 2.76~TeV~\cite{Aad:2015lwa}. The experimental data from ATLAS are shown by the closed circles. The results simulated by AMPT and HYDJET++ models are shown by the open red triangles and blue squares, respectively. The shadow boxes represent the systematic uncertainties of the experimental data, while the statistical uncertainties are smaller than the symbol size.}
\end{figure}

Centrality dependencies of the Fourier harmonics $v_{2}$, $v_{3}$ and $v_{4}$ from PbPb collisions at $\sqrt{s_{_{\mathrm{NN}}}}=2.76$~TeV simulated by the HYDJET++ and AMPT models are shown in Fig.~\ref{fig:1} together with the experimental data from the ATLAS Collaboration~\cite{Aad:2015lwa}. In both models, the elliptic $v_{2}$ harmonic exhibits a strong centrality dependence, while the $v_{3}$ and $v_{4}$ have a weak centrality dependence. The $v_{2}$ in the HYDJET++ model continuously increases up to 70\% centrality, while in the AMPT model it reaches its maximum at 50--60\% centrality and than starts to decrease similarly to the experimental data. Except the most central collisions where predictions from both models agree with the experimentally measured $v_{2}$, the models predict smaller elliptic flow than the data for centralities up to 50\%. For centralities above 50\%, the $v_{2}$ from the HYDJET++ model continues to increase and becomes greater than the expermental one. The $v_{3}$ extracted from both models is, up to 50\% centrality, in a mutual agreement and in a good agreement with the experimentally measured $v_{3}$. For centralities above 50\%, both models give $v_{3}$ greater than the experimentally measured one. The AMPT model prediction for $v_{4}$ Fourier harmonic is in a very good agreement with the experimentally measured $v_{4}$ for practically the whole centrality range, while the corresponding prediction from the HYDJET++ model disagrees with the experimental data. We note that the fourth-order spatial and momentum anisotropies are not implemented in the HYDJET++ model so the failure of the model in $v_{4}$ is expected. Based on these calculations, the correlations between Fourier harmonics of different orders are presented in the rest of the paper.

\begin{figure}
  \includegraphics[width=0.7\textwidth]{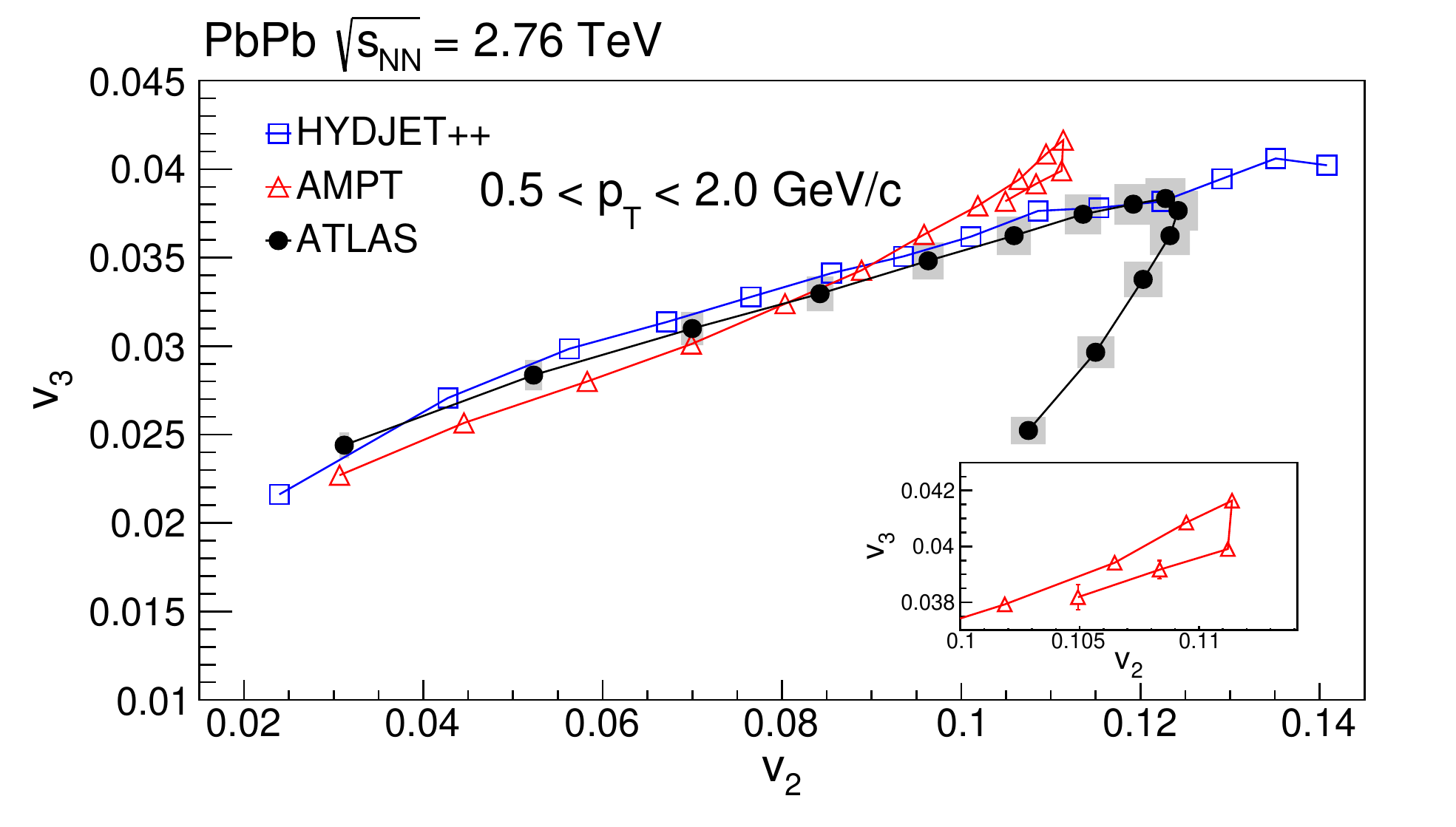}
  \caption{\label{fig:2} (Color online) The correlation between $v_{2}$ and $v_{3}$ from the 0.5 $< p_{T} <$ 2~GeV/c interval for fourteen 5\%-wide centrality classes over the centrality range 0--70\% in PbPb collisions at 2.76~TeV~\cite{Aad:2015lwa}. The results simulated by the AMPT and HYDJET++ models are shown by the open red triangles and blue squares, respectively. The shadow boxes represent the systematic uncertainties of the experimental data, while the statistical uncertainties are smaller than the symbol size.}
\end{figure}

The correlation between the average $v_{2}$ and $v_{3}$ Fourier harmonics, where each point represents one centrality class, is shown in Fig.~\ref{fig:2}. In contrast to the $v_2$, higher order Fourier harmonics $v_{n}$ ($n=3,4$) have a weak centrality dependence and in peripheral collisions they decrease faster than in central collisions as measured by ATLAS~~\cite{Aad:2015lwa}. This introduces the appearance of a boomerang-like structure. The ideal hydrodynamics in the HYDJET++ model predicts nearly linear centrality dependence of the $v_{n}$ harmonics, and thus does not produce the boomerang structure. The same behavior has been found in~\cite{Qian:2016pau} with specific shear viscosity $\eta/s$ = 0, while in the viscous hydrodynamics with positive $\eta/s$ values, the $v_{n}$ harmonics reach maxima and then decrease going to peripheral collisions. As a consequence, the boomerang-like shape appears in $v_{n}$--$v_{m}$ correlations. The slopes of the $v_{2}$--$v_{3}$ dependence agree between the experimental data and the HYDJET++ model up to about 50\% centrality. Going to more peripheral collisions, the difference starts to increase and show qualitatively different behavior. Up to 50\% centrality, the magnitudes of the $v_{2}$ coefficients in the experimental data are greater than the ones predicted in the HYDJET++ model. For centralities above 50\%, HYDJET++ predicts further increases of the $v_{2}$ and $v_{3}$ coefficients, while the experimental data show a decrease.

Unlike HYDJET++, the AMPT model predicts the slope as well as the boomerang-like shape of the correlation. However, the boomerang turn is much sharper than the one seen in the ATLAS data~\cite{Aad:2015lwa} (see the zoomed plot in the Fig.~\ref{fig:2}). The experimental data from peripheral collisions shows a much faster decrease of the $v_{3}$ coefficient than the AMPT prediction where the decrease of the $v_{3}$ coefficient going to the peripheral collisions is the same as the decrease going to central collisions. For centralities above 35\%, the $v_{3}$ coefficients from the AMPT model are greater than the ones measured by ATLAS, while for centralities below 20\% the model results are somewhat smaller than the data. The relatively small partonic cross section used for this analysis of the AMPT model induces a significant effective specific shear viscosity~\cite{Zhou:2016eiz}, which produces maxima in the $v_{n}$ distributions and then a decrease of the $v_{n}$ going towards peripheral collisions.
  
\begin{figure}
  \includegraphics[width=0.7\textwidth]{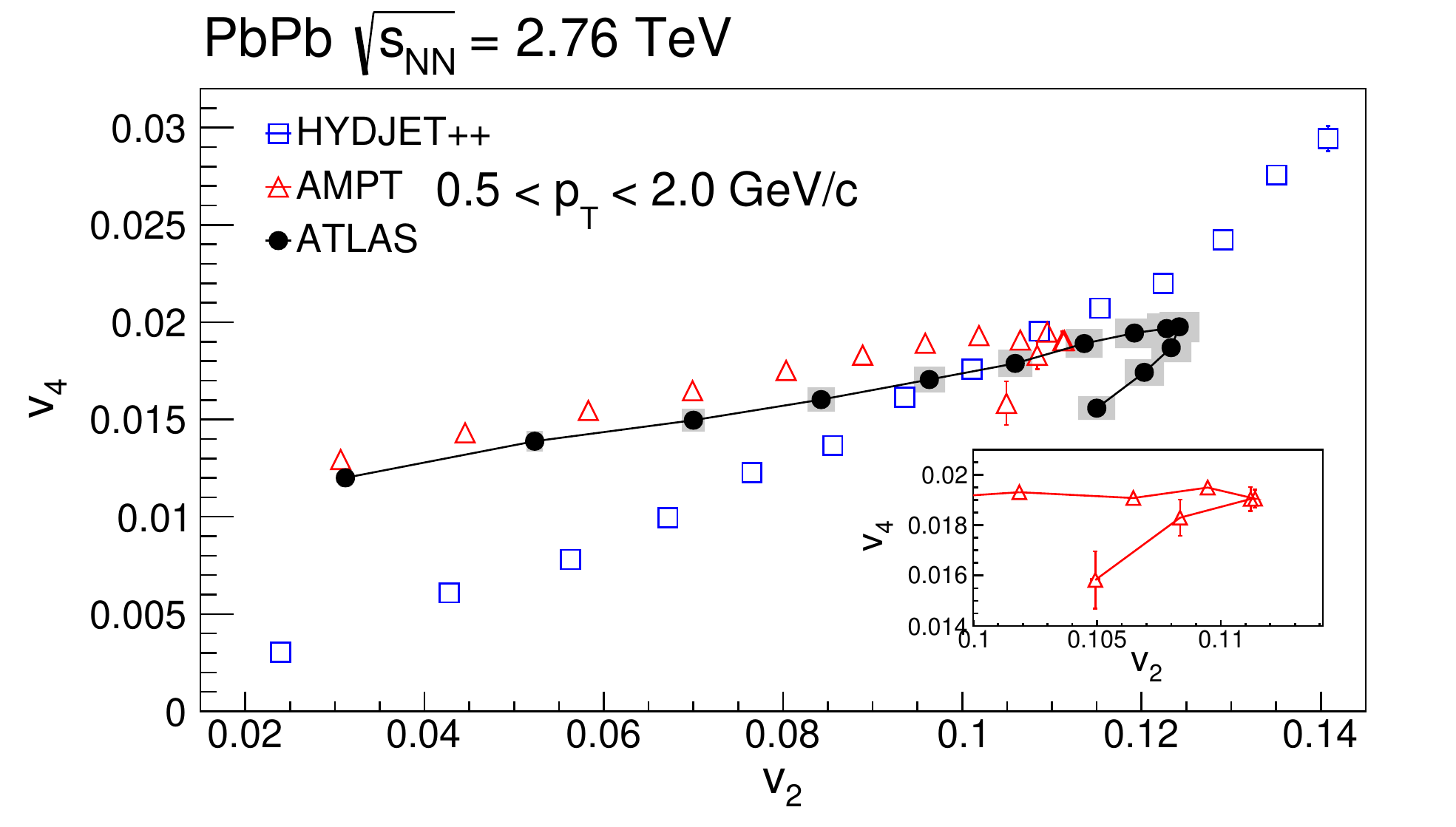}
\caption{\label{fig:3} (Color online) The correlation between $v_{2}$ and $v_{4}$ from the 0.5 $< p_{T} <$ 2~GeV/c interval for thirteen (fourteen) 5\%-wide centrality classes over the centrality ranges 0--65\% (0--70\%) in PbPb collisions at 2.76~TeV~\cite{Aad:2015lwa}. The results simulated by the AMPT and HYDJET++ models are shown by the open red triangles and blue squares, respectively. The error bars represent the statistical uncertainties. The shadow boxes represent the systematic uncertainties of the experimental data.}
\end{figure}

Fig.~\ref{fig:3} shows centrality dependence of the correlation between $v_{2}$ and $v_{4}$ harmonics. Again, the experimental data~\cite{Aad:2015lwa} for very peripheral collisions show a boomerang shape, while the HYDJET++ model does not predict it. The HYDJET++ model predicts a much stronger slope than the one seen in the experimental data. Also, in central collisions, the HYDJET++ model predicts smaller $v_{4}$ values, while in peripheral collisions HYDJET++ gives greater $v_{4}$ values than the ones seen in the experimental data. Experimental $v_{2}$ values are greater than those extracted from the HYDJET++ simulation except for centralities above 50\% where $v_{2}$ values continue to increase, while experimental ones start to decrease. In contrast to the $v_{3}$ anisotropy, the $v_{4}$ anisotropy is not introduced in the model via the corresponding eccentricity. Appearance of the higher order harmonics $v_{n}$ ($n > $3) is a consequence of the interference of the $v_{2}$ and $v_{3}$ harmonics~\cite{Bravina:2013xla}. As there is a larger disagreement between HYDJET++ $v_{4}$ predictions and the data, this model does not describe the data well also in the $v_{2}$--$v_{4}$ correlation analysis.

Fig.~\ref{fig:3} also shows prediction of the AMPT model. The AMPT model reproduces the slope and the boomerang-like shape of the $v_{2}$--$v_{4}$ correlation (see the zoomed plot in the Fig.~\ref{fig:3}). In contrast to the $v_{2}$--$v_{3}$ correlation, the AMPT model predicts the experimentally observed slope of the $v_{2}$--$v_{4}$ correlation even for peripheral collisions. The model reproduces rather well the experimentally measured $v_{4}$ values, while the $v_{2}$ values from AMPT are smaller than the measured ones except for the most central 0--5\% collisions.

\begin{figure}
  \includegraphics[width=0.7\textwidth]{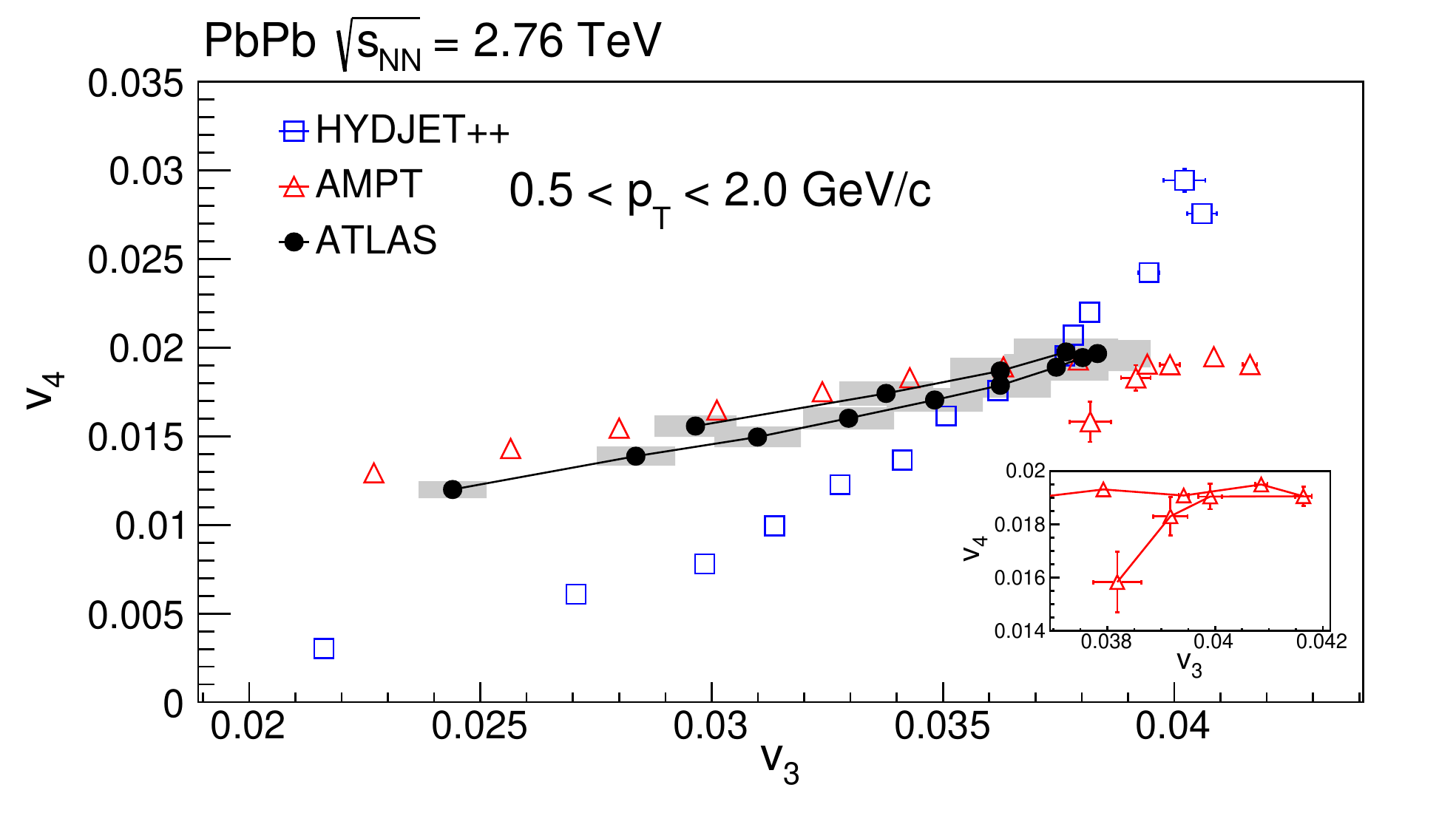}
  \caption{\label{fig:4} (Color online) The correlation between $v_{3}$ and $v_{4}$ from the 0.5 $< p_{T} <$ 2~GeV/c interval for thirteen (fourteen) 5\%-wide centrality classes over the centrality range 0--65\% (0--70\%) in PbPb collisions at 2.76~TeV~\cite{Aad:2015lwa}. The results simulated by the AMPT and HYDJET++ are shown by the open red triangles and blue squares, respectively. The error bars represent the statistical uncertainties. The shadow boxes represent the systematic uncertainties of the experimental data.}
\end{figure}

The correlation between $v_{3}$ and $v_{4}$ Fourier harmonics is shown in Fig.~\ref{fig:4}. Again, due to a larger disagreement between HYDJET++ $v_{4}$ predictions and the data, the HYDJET++ model predicts a steeper slope of $v_{3}$--$v_{4}$ correlation than the experimental data~\cite{Aad:2015lwa}. Similarly to the case of the $v_{2}$--$v_{4}$ correlation, the AMPT model reproduces the $v_{3}$--$v_{4}$ correlation observed in the experiment. For the peripheral collisions the AMPT model predicts the boomerang-like shape of the correlations with an opening angle similar to the one seen in the experimental data. However, it is worthwhile to note that there seems a subtle difference, namely, the data show a boomerang-like shape with a left turn, while the AMPT model predicts a right turn. Except for the most peripheral collisions, the AMPT model reproduces the magnitudes of the $v_{4}$ Fourier harmonic very well. For central collisions, the $v_{3}$ magnitudes are somewhat smaller than the experimentally  measured ones, while for peripheral collisions the situation is opposite.

\section{Summary}
\label{sec:conc}
Centrality dependence of the correlations between $v_{2}$ and $v_{3}$, $v_{2}$ and $v_{4}$, and $v_{3}$ and $v_{4}$ are studied using two-particle correlation technique within 0.5~$< p_{T} <$~2.0~GeV/c and $|\eta| <$~2.5 in PbPb collisions at $\sqrt{s_{_{\mathrm{NN}}}}=2.76$~TeV simulated by the HYDJET++ and AMPT models. The results are compared to the corresponding experimental measurements obtained by the ATLAS Collaboration. In general, both models reproduce rather well the experimentally measured magnitudes of the Fourier harmonics $v_{n}$ in central collisions. Going to more peripheral collisions, discrepancy between data and models becomes more pronounced. In the case of the correlations between $v_{2}$ and $v_{3}$, both the HYDJET++ and AMPT models reproduce rather well the slope. Because of a weak centrality dependence of the higher-order $v_{3}$ and $v_{4}$ Fourier harmonics, and a faster decrease of their magnitudes in peripheral than in central collisions, the experimental data exhibit a boomerang-like shape. This structure is not observed in the HYDJET++ model, but is reproduced by the AMPT model. Due to the disagreement between HYDJET++ $v_{4}$ predictions and the data, the HYDJET++ model does not reproduce the slopes in the correlations between $v_{2}$ and $v_{4}$ and between $v_{3}$ and $v_{4}$, while the AMPT model reproduces them well. In conclusion, the AMPT model reproduces the experimentally observed features of the correlations between Fourier harmonics of different orders better than the HYDJET++ model.

\begin{acknowledgments}
The authors acknowledge the support of the Bilateral Cooperation 
between Republic of Serbia and People's Republic of China 451-03-478/2018-09/04 ''Phenomenology in high energy physics''. This work was supported in part by the Ministry of Education, Science and Technological Development of the Republic of Serbia (Grant No.~171019), the National Natural Science Foundation of China (Grant No.~11847315) and the U.S.~Department of Energy (Grant No.~de-sc0012910).
\end{acknowledgments}

\end{document}